\documentclass[%
 preprint,
superscriptaddress,
 amsmath,amssymb,
 aps,
prl,
]{revtex4-1}

\usepackage{graphicx}
\usepackage{dcolumn}
\usepackage{bm}
\usepackage{amsmath}
\usepackage{color}
\usepackage[colorlinks=true,bookmarks=false,citecolor=blue,urlcolor=blue]{hyperref}

\graphicspath{ {./Figures/} }

\newcommand{\eps}{\varepsilon}

\begin{document}


\title{Anapole states and scattering deflection effects in anisotropic van der Waals nanoparticles}

\author{Andrei A. Ushkov}
\email{ushkov.aa@mipt.ru}
\affiliation{Center for Photonics \& 2D Materials, Moscow Institute of Physics and Technology (MIPT), Dolgoprudny 141700, Russia}
\affiliation{Department of Electrical Engineering Tel Aviv University Ramat Aviv, Tel Aviv 69978, Israel}
 
\author{Georgy A. Ermolaev}%
\affiliation{Center for Photonics \& 2D Materials, Moscow Institute of Physics and Technology (MIPT), Dolgoprudny 141700, Russia}

\author{Andrey A. Vyshnevyy}
\affiliation{Center for Photonics \& 2D Materials, Moscow Institute of Physics and Technology (MIPT), Dolgoprudny 141700, Russia}

\author{Denis G. Baranov}
\affiliation{Center for Photonics \& 2D Materials, Moscow Institute of Physics and Technology (MIPT), Dolgoprudny 141700, Russia}

\author{Aleksey V. Arsenin}
\affiliation{Center for Photonics \& 2D Materials, Moscow Institute of Physics and Technology (MIPT), Dolgoprudny 141700, Russia}

\author{Valentyn S. Volkov}
\affiliation{Center for Photonics \& 2D Materials, Moscow Institute of Physics and Technology (MIPT), Dolgoprudny 141700, Russia}

\date{\today}

\begin{abstract}
Transition metal dichalcogenides (TMDCs), belonging to the class of van der Waals materials, are promising materials for optoelectronics and photonics. In particular, their giant optical anisotropy may enable important optical effects when employed in nanostructures with finite thickness. In this paper, we theoretically and numerically study light scattering behavior from anisotropic MoS$_2$ nanocylinders, and highlight its distinct features advantageous over the response of conventional silicon particles of the same shape. We establish two remarkable phenomena, appearing in the same MoS$_2$ particle with optimized geometry. The first one is a pure magnetic dipole scattering associated with the excitation of the electric-dipole anapole states. Previously reported in core-shell hybrid (metal/dielectric) systems only, it is now demonstrated in an all-dielectric particle. The second phenomenon is the super-deflection in the far field: the maximum scattering may occur over a wide range of directions, including forward-, backward- and side-scattering depending on the mutual orientation of the MoS$_2$ nanocylinder and the incident wave. In contrast to the well-known Kerker and anti-Kerker effects, which appear in nanoparticles at different frequencies, the super-deflection can be achieved by rotating the particle at a constant frequency of incident light. Our results facilitate the development of functional optical devices incorporating nanostructured anisotropic TMDCs and may encourage further research in meta-optics based on highly anisotropic materials.
\end{abstract}
  
\maketitle

\section{Introduction}

Light manipulation by resonant optical nanoparticles made from high-index dielectrics \cite{evlyukhin2012demonstration} is an entire branch of nanophotonics \cite{kuznetsov2016optically} with a wide spectrum of applications ranging from flat optics \cite{genevet2017recent,babicheva2021multipole} to bio-spectroscopy \cite{krasnok2018spectroscopy}.
Thanks to the rich variety of multipolar modes supported by such nanoparticles \cite{terekhov2017multipolar}, they exhibit unique interference effects, such as tunable scattering patterns \cite{fu2013directional},  Kerker/anti-Kerker effects \cite{person2013demonstration,shamkhi2019transverse}, nonlinear harmonic generation \cite{smirnova2016multipolar}, and many others \cite{koshelev2020dielectric}. Due to their subwavelength dimensions, nanoparticles are perfect for dense integration into functional optical devices without sufficient increase of the device dimensions.
Traditional material platforms for the implementation of resonant dielectric nanoparticles include bulk semiconductors, such as Si, GaAs, Ge, and others \cite{baranov2017all}.
Recently, transition metal dichalcogenides (TMDCs) -- a family of van der Waals (vdW) materials, have been recognized as a novel promising platform for dielectric nanostructures.

VdW materials have been a subject of intense research for more than a decade and have demonstrated unique electronic and optical properties \cite{Mak2016, wang2020exciton}.
Although these materials are often employed in their two-dimensional form, vdW materials also allow fabrication of various nanostructures with wavelength-scale thickness \cite{munkhbat2018self,ling2021all,verre2019transition}.
The highly anisotropic atomic/crystal lattice of the vdW materials family causes prominent anisotropy in the optical response. Both the out-of-plane ($\eps_{xx} = \eps_{yy} \ne \eps_{zz}$) \cite{ermolaev2020broadband} and in-plane anisotropy ($\eps_{xx} \ne \eps_{yy}$) \cite{mao2016optical,gogna2020self} is possible.
The anisotropy of TMDCs in the visible and near-infrared spectral ranges presents a new degree of freedom for nanoparticle design. 
In the context of light scattering by compact nanoparticles, it is particularly interesting to demonstrate novel functionalities enabled by the optical anisotropy.

A number of recent works have been devoted to both theoretical and experimental exploration of TMDC-based nanoresonators.
Studies have already utilized this anisotropy for probing internal fields of TMDC nanoantennas via anisotropic Raman scattering \cite{green2020optical}, excitation of in-plane anisotropic exciton-polaritons \cite{gogna2020self}, and reducing the cross-talk effect between parallel waveguides \cite{ermolaev2021giant}.
The ability of TMDCs to form nanostructures with sharp edges is also attractive for optoelectronics and catalysis \cite{munkhbat2020transition}.
Besides linear optical effects, nanoparticles made of TMDCs have also shown unexpected regimes of second harmonic generation \cite{busschaert2020transition,tselikov2021double,nauman2021tunable}.

In this paper, we investigate theoretically and numerically the light scattering from nanocylinders made of MoS$_2$ compared to analogous structures made of silicon, the most common material in photonics nowadays. In particular, we show that due to specific MoS$_2$ excitonic band a frequency-selective transverse scattering appears in nanocylinder, which is difficult or even impossible to reproduce with conventional isotropic materials and particles. Furthermore, we study the beam-steering capabilities of MoS$_2$ nanocylinders in visible and near-infrared and reveal an interesting ``super deflector" regime, when the deflection can be tuned in the full angular range from 0$^\circ$ (forward scattering) to 180$^\circ$ (backward scattering), including all intermediate directions, by rotation of the nanoparticle. We expect that the obtained results will be useful for meta-optics of highly anisotropic particles, all-dielectric photonic device engineering, and will encourage further fundamental and applied research of TDMCs in optics.

\section{Anapole states in Si and MoS$_2$ nanoparticles}

\subsection{Particle geometry and simulation setup}
\begin{figure*}[h!]
\centering\includegraphics[width=16 cm]{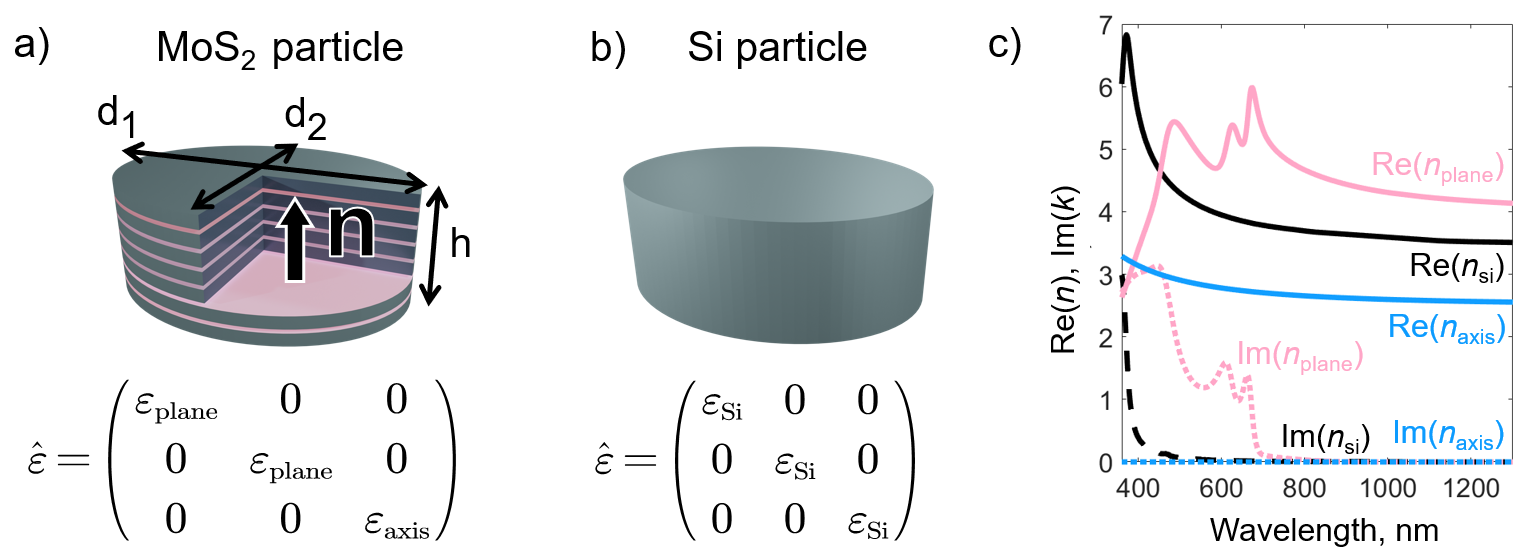}
\caption{a)-b) Schematic view of the studied elliptic-cylinder nanoparticles. A free-standing MoS$_2$/Si nanoparticle in vacuum is illuminated with TE/TM polarized light in visible and near-IR; the TE/TM polarization is defined with respect to the incidence plane formed by the normal vector \textbf{n} and $d_1$. Optic axis of MoS$_2$ is oriented along the vector \textbf{n}. c) Optical constants of MoS$_2$ and Si used for numerical simulations.}
\label{MoS2andSilicon.PNG}
\end{figure*}

Figure \ref{MoS2andSilicon.PNG}a schematically shows the studied nanostructure. We consider a cylindrical MoS$_2$ (and Si) nanoparticle as a standard experimentally feasible geometry, with a height $h$ and an elliptic base with principal axes $d_1$ and $d_2$. In contrast to silicon (Fig. \ref{MoS2andSilicon.PNG}b), MoS$_2$ is a highly anisotropic negative uniaxial crystal with its in-plane refractive index being $\sim60\%$ larger than the out-of-plane index, see Fig. \ref{MoS2andSilicon.PNG}c. To emphasize this visually, a MoS$_2$ particle is illustrated as a layered structure in Fig. \ref{MoS2andSilicon.PNG}a; vector \textbf{n} indicates the optic axis. In the following, the incident light wave vector \textbf{k} always lies in the plane formed by $d_1$ and \textbf{n}.


Numerical simulations of light scattering by the MoS$_2$ nanoparticle were performed using commercial finite-difference time-domain software (Lumerical). We used the experimentally measured optical properties from ref. \cite{ermolaev2021giant} and Palik’s handbook \cite{palik1998handbook} for MoS$_2$ and silicon, respectively. 
The multipole analysis of scattered light was performed following the exact integral multipole decomposition approach \cite{alaee2019exact}:
\begin{multline}
\left[ \begin{array}{c}
	q_{jm}^{E}\\
	q_{jm}^{M}\\
\end{array} \right] =-\left( i \right) ^j\frac{\sqrt{\mu_0\mu_d /\varepsilon_0\varepsilon_d}k_d^2}{2}\frac{4\pi}{\sqrt{\left( 2\pi \right) ^3}}\cdot 
\\
\sum_{\bar{l},\bar{m}}^{}{\left( -i \right) ^{\bar{l}}\int{d\mathbf{\hat{p}}\left[ \begin{array}{c}
	\mathbf{Z}_{j,m}^{\dagger}\left( \mathbf{\hat{p}} \right)\\
	\mathbf{X}_{j,m}^{\dagger}\left( \mathbf{\hat{p}} \right)\\
\end{array} \right] Y_{\bar{l}\bar{m}}\left( \mathbf{\hat{p}} \right)}\int{d^3\mathbf{r}\,\,\mathbf{J}\left( \mathbf{r} \right) Y_{\bar{l}\bar{m}}^{*}\left( \mathbf{\hat{r}} \right) j_{\bar{l}}\left( k_dr \right)}}
,
\label{Eq1}
\end{multline}
where $\mathbf{J}\left( \mathbf{r} \right) =-i\omega \left( \mathbf{r} \right) \left[ \hat{\eps} _{\rm particle}\left( \mathbf{r} \right) - \eps_d \right]\varepsilon_0\mathbf{E}$ is the induced electric current density within the particle with the dielectric permittivity tensor $\hat{\eps} _{\rm particle}$; the relative permittivity and permeability of the embedding medium is $\eps _{d}$ and $\mu_{d}$, respectively; $k_d$ is the wavevector of the incident light in embedding medium. 
In the above equation, $q_{jm}^{E}$ and $q_{jm}^{M}$ correspond to electric and magnetic dipoles of scattering field ($j=1$), quadrupoles ($j=2$) etc, respectively; $\bar{l}=j\pm1$ for electric and $\bar{l}=j$ for magnetic multipoles, respectively; $\bar{m}=-\bar{l}...\bar{l}$. Functions $\mathbf{Z}_{j,m}$ and $\mathbf{X}_{j,m}$ are the eigenstates of operators of the squared angular momentum and the $z$-component of angular momentum; $Y_{jm}\left( \mathbf{\hat{r}} \right)$ are the spherical harmonics and $j_l(x)$ are spherical Bessel functions; $\mathbf{\hat{r}}\equiv \mathbf{r}/r$ and $\mathbf{\hat{p}}\equiv \mathbf{p}/p$ are unit vectors in the direction of ${\bf r}$ and ${\bf p}$, respectively; the first integral on the right side of Eq. \ref{Eq1} is over the unit sphere in reciprocal space, the second integral is over the particle volume in a real space. The scattering cross-section $\sigma_{\rm sca}$ is calculated via \cite{alaee2019exact}:
\begin{equation}
\sigma _{\rm sca}=\frac{1}{k_d^2}\sum_{j>0}^{}{\sum_{m=-j}^j{\left| q_{jm}^{E} \right|^2+\left| q_{jm}^{M} \right|^2}}
\label{Eq2}
\end{equation}
 Exact dipole Cartesian components $\mathbf{p}, \mathbf{m}$ were calculated via equations developed in \cite{evlyukhin2019multipole} and were used for the dipole approximation and visualization. The particle scattering cross-section in dipole approximation using exact Cartesian dipoles representation reduces to \cite{evlyukhin2019multipole}:

\begin{equation}
    \sigma _{\rm sca}\simeq \frac{k_{0}^{4}}{12\pi \varepsilon _{0}^{2}v_d\mu _0}\left| \mathbf{p} \right|^2+\frac{k_{0}^{4}\varepsilon _d}{12\pi \varepsilon _0v_d}\left| \mathbf{m} \right|^2,
\label{Eq3}
\end{equation}
where $k_{0}$ is the wavenumber in vacuum, $v_d$ is the speed of light in the embedding medium. 

\subsection{Results}

\begin{figure*}
\centering\includegraphics[width = \textwidth]{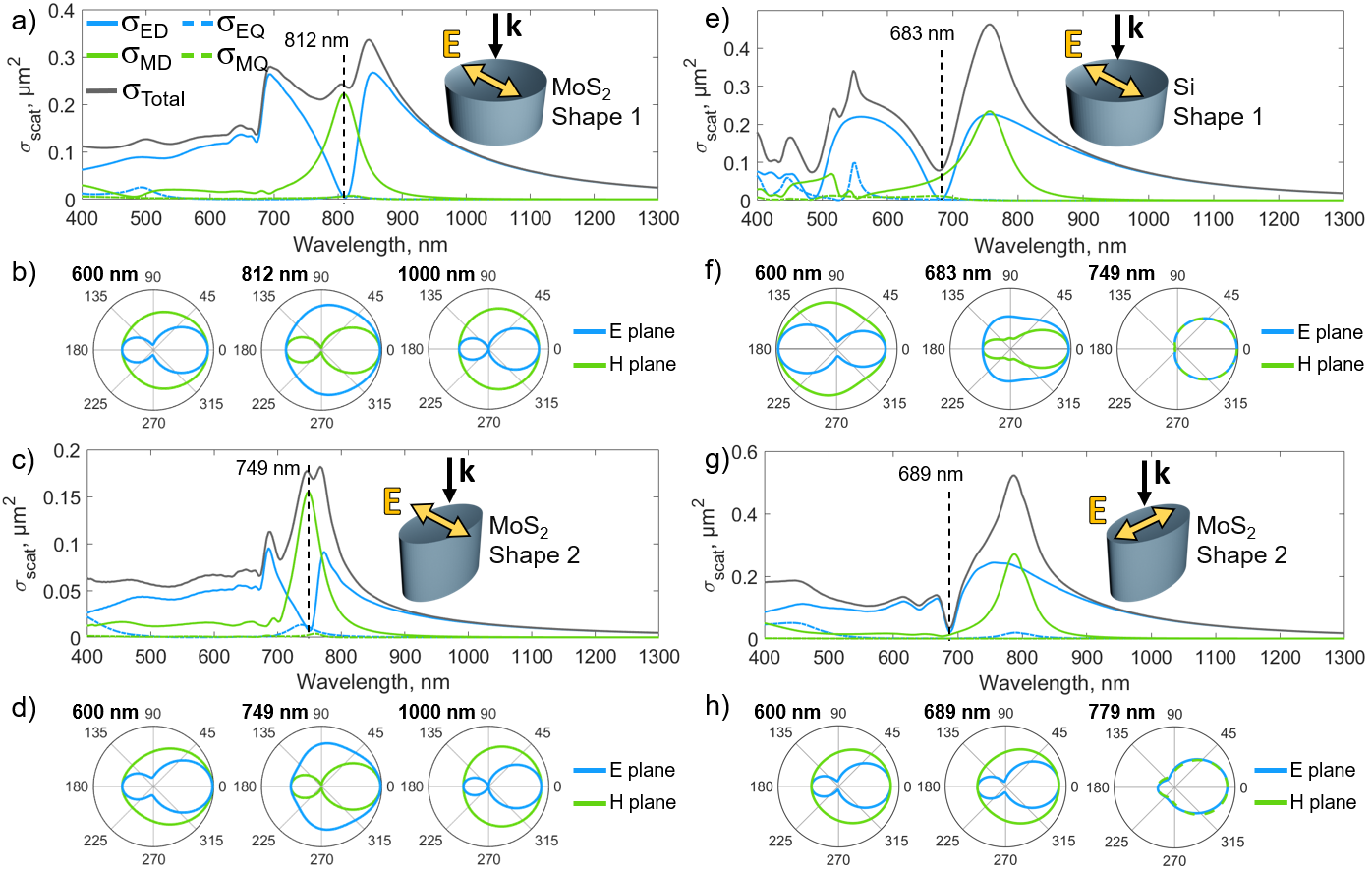}
\caption{Spectral analysis of anapoles and magnetic dipoles in MoS$_2$ and Si particles. (a) Scattering multipole decomposition of MoS$_2$ particle with shape 1 with $d_1=d_2=260$~nm, $h=120$~nm (see Fig. \ref{MoS2andSilicon.PNG}a for geometric parameters). The magnetic scattering occurs at the wavelength 812~nm. Incident light is a linearly polarized plane wave falling vertically on the particle with respect to the inset in a); b) Far field scattering patterns at wavelengths 600~nm, 812~nm and 1000~nm from left to right, in two mutually orthogonal planes: E plane contains the wave vector and electric field amplitude of the incident light, H plane contains the wave vector and magnetic field amplitude of the incident light. c) and d) are analogous to a) and b) for the MoS$_2$ particle shape 2 with $d_1=260$~nm, $d_2=140$~nm, $h=150$~nm. Electric field of the incident light is oriented along the short axis of the particle ellipse, as shown by orange arrow in the inset; e) and g) present scattering multipole decomposition for particles made of Si and MoS$_2$, with shapes 1 and 2, respectively; f) and h) are the corresponding far field scattering patterns at characteristic wavelengths.}
\label{MultipolesScatteringAndField.PNG}
\end{figure*}

We start our analysis of TMDC-based nanoresonators with the demonstration of an ideal magnetic dipole scattering. Figure \ref{MultipolesScatteringAndField.PNG}a shows the total scattering cross-section of the MoS$_2$ nanodisk in visible and near-IR, together with its multipole components up to quadrupoles. The particle is a circular cylinder with optimized geometrical parameters $d_1=d_2=260$~nm, $h=120$~nm, the incident light wave vector is oriented along the vector $\mathbf{n}$ (see Fig. \ref{MoS2andSilicon.PNG}a for geometric definitions). The incident plane wave is polarized linearly, as it is shown in graphical insets in Figs. \ref{MultipolesScatteringAndField.PNG}a,c,e,g.
In the following, we refer to this particular geometry of the nanocylinder with  $d_1/h=d_2/h \approx 2.17$ as ``shape 1", see the inset in Fig. \ref{MultipolesScatteringAndField.PNG}a. The particle shape was optimized to match an anapole in the electric dipole channel with a resonance in the magnetic dipole channel at the wavelength of $\lambda = 812$~nm in a region with near-zero excitonic absorption (see Fig. \ref{MoS2andSilicon.PNG}c). With the higher-order multipoles sufficiently suppressed and the electric dipole scattering eliminated, the magnetic dipole contribution to scattering becomes dominant at $\lambda=812$~nm. Far-field scattering patterns (Fig. \ref{MultipolesScatteringAndField.PNG}b) also show dominant magnetic dipole scattering at $\lambda=812$~nm.

In the following, we demonstrate that the ideal magnetic dipole scattering effect first introduced in ref. \cite{feng2017ideal} for Au core/silicon shell nanoparticles can be reproduced using a homogeneous non-magnetic material as well. Moreover, the MoS$_2$ particle shape can be further optimized to enhance the magnetic dipole response. 

Figure \ref{MultipolesScatteringAndField.PNG}c presents the multipole expansion of scattered field by a MoS$_2$ particle shaped as an elongated elliptical cylinder with dimensions $d_1=260$~nm, $d_2=140$ nm, $h=150$~nm. We refer to this particular geometry with $d_1/h\approx1.73$, $d_2/h\approx 0.93$ as ``shape 2" (see the inset in Fig. \ref{MultipolesScatteringAndField.PNG}c). The incident plane wave is oriented similarly to the case of Fig. \ref{MultipolesScatteringAndField.PNG}a, with the electric field polarized along the minor axis $d_2$. The particle has the same optical regime of anapole and magnetic dipole resonance overlapping as in Fig. \ref{MultipolesScatteringAndField.PNG}a, but additionally, in contrast, demonstrates the magnetic dipole resonant scattering at $\lambda=749$~nm higher than one coming from all other scattering channels in visible and near-IR bands. The far-field scattering pattern at $\lambda=749$~nm in Fig. \ref{MultipolesScatteringAndField.PNG}d generally repeats those from Fig. \ref{MultipolesScatteringAndField.PNG}b; slight variations are caused by the increased contribution of higher-order multipoles to the scattering.

Dielectric materials conventionally used in optics and photonics (silicon, glass, titanium dioxide and others) are typically required to exhibit low absorption at operating wavelengths, as it leads to heating and generally limits the optical response of the nanostructure \cite{khurgin2015deal, baranov2017all}. In the present case, however, absorption in MoS$_2$ at wavelengths shorter than 700 nm favours the ideal magnetic dipole scattering regime in MoS$_2$ nanodisks. 
Indeed, the broad exciton band in MoS$_2$ below 700~nm (see Fig. \ref{MoS2andSilicon.PNG}c) suppresses Mie resonances in this spectral region, and renders the particle an over-damped electric dipole scatterer (Figs. \ref{MultipolesScatteringAndField.PNG}a,c). 
At wavelengths above $\sim 900$ nm, on the other hand, the nanoparticle  does not support any Mie resonances due to its insufficient dimensions and also behaves as a non-resonant electric dipole scatterer. 
Therefore, MoS$_2$ offers a particle design with a relatively narrow window of resonant behavior (700-900~nm). In this work we use this window for anapole state-magnetic dipole resonance overlap, and the mechanisms described above ensure this resonance to be the only one in visible-near IR frequency range.

Another remarkable feature of the ``shape 2" MoS$_2$ particle design is its sensitivity to the polarization state of incident light. The polarization switching at 749~nm initiates the transition from the strong magnetic scattering in Fig.~\ref{MultipolesScatteringAndField.PNG}c to the mixed scattering regime with both electric and magnetic dipole components in Fig.~\ref{MultipolesScatteringAndField.PNG}g. Such kind of transition is challenging to reproduce in particles made of isotropic materials (such as silicon), as they have only a geometric degree of freedom for the anapole-dipole overlay optimization design.

\begin{figure}
\centering\includegraphics[width =.7\textwidth]{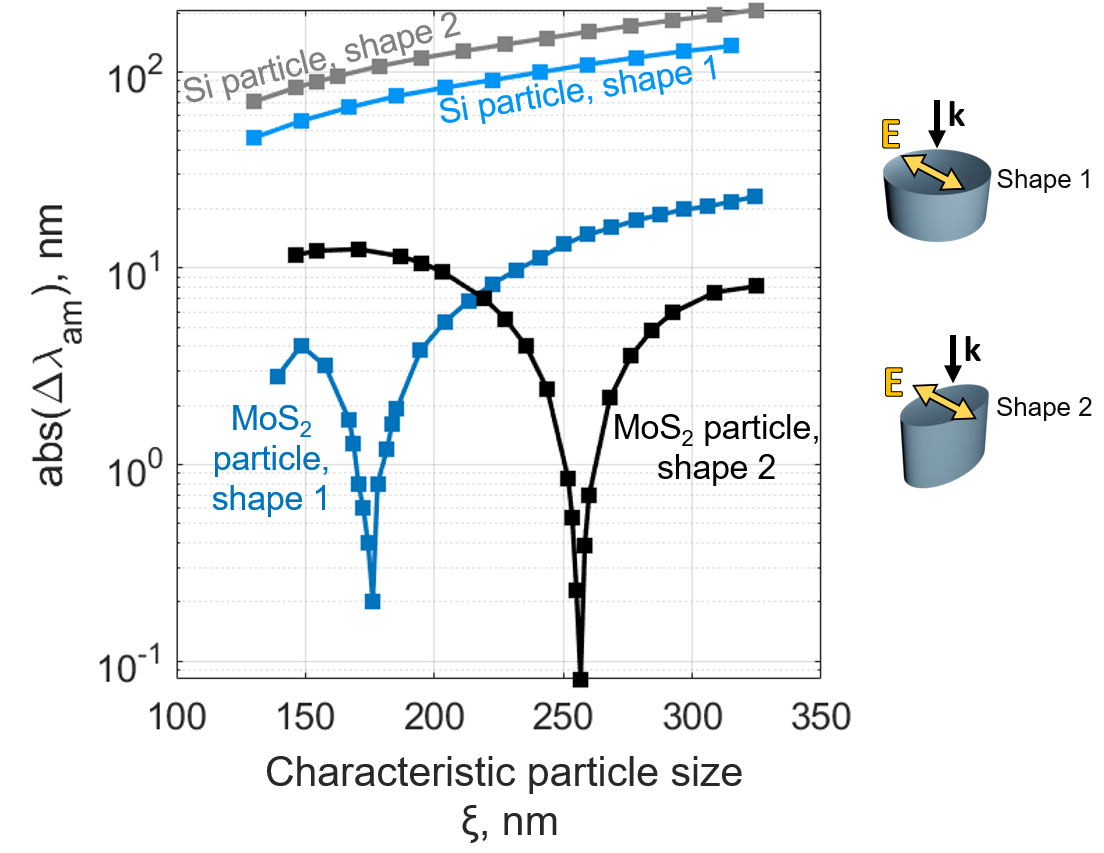}
\caption{The spectral distance $|\Delta\lambda_{am}|$ between the anapole in electric dipole channel and the magnetic dipole resonance as a function of a characteristic particle size $\xi=\sqrt[3]{V_{\rm particle}}$. Each curve corresponds to the particle of certain shape and material, denoted by the label near the line. The incident light direction and polarization with respect to the particles with shapes 1 and 2 are denoted to the right of the graph.}
\label{AnapoleDipoleShift.PNG}
\end{figure}

The anisotropy and strong frequency dispersion of the MoS$_2$ permittivity tensor facilitates the design of ideal magnetic resonant scatterers. 
Scaling the particle's volume made of a dispersion-less material while preserving its shape shifts the wavelengths of its electric and magnetic resonances proportionally, but keeps their relative spectral order unchanged \cite{evlyukhin2010optical,feng2017ideal}. 
Scaling the volume of the silicon particles (and the corresponding change of characteristic particle size $\xi=\sqrt[3]{V_{\rm particle}}$) changes the spectral distance $\Delta\lambda_{am} = |\lambda_{a} - \lambda_{m}|$ between the anapole and the magnetic dipole resonance (see Fig. \ref{AnapoleDipoleShift.PNG}), but do not swap them spectrally, despite the presence of frequency dispersion in Si permittivity.

In contrast, the currents induced by incident light in MoS$_2$ particles are more strongly affected by the dispersion and anisotropy of the MoS$_2$ permittivity tensor. 
Whereas the electric dipole moment (which, in the present configuration, is aligned with the incident electric field) is formed mainly by currents parallel the MoS$_2$ crystal planes, the magnetic dipole is determined by currents perpendicular to these planes for the particular orientation of the incident plane wave as shown in Fig.~\ref{AnapoleDipoleShift.PNG}. 
Consequently, the electric and magnetic dipole moments induced in the nanoparticle ``feel" different frequency dispersions of the in- and out-of-plane MoS$_2$ permittivity tensor components.
This leads to a different rate of spectral red-shift of the induced anapole $\lambda_{a}(\xi)$ in the electric dipole channel and magnetic dipole resonance $\lambda_{m}(\xi)$ with the increase of the particle size $\xi$ (while preserving its shape).
As a result of these different rates, the distance $\Delta\lambda_{am}$ may approach zero for the case of MoS$_2$ nanoparticle. 
Figure \ref{AnapoleDipoleShift.PNG} clearly shows zeros of $|\Delta\lambda_{am}|$ for MoS$_2$ particles of shapes 1 and 2. 
The material anisotropy provides a new degree of freedom for particle design which provides the anapole-magnetic resonance overlap.

\section{Kerker regimes in asymmetric MoS$_2$ nanoparticles}



One of the salient features of dielectric nanoparticles that determined their wide range of applications in nanophotonics is the capability to host simultaneously electric and magnetic multipolar resonances of various order, amplitude and relative phase. The interference between different multipoles excited in the nanoparticle produces complicated far-field scattering patterns evolving with wavelength.
This section is devoted to analysis of the generalized Kerker regimes in MoS$_2$ nanoparticles. 
We examine the effect of the particle rotation on the scattering pattern at the fixed wavelength of 818~nm, which is nearly in the center of spectral window 700-900~nm with a resonant Mie behaviour (see Figs.\ref{MultipolesScatteringAndField.PNG}a,c). The study is limited to a single particle ``shape 2" only with dimensions from Fig. \ref{MultipolesScatteringAndField.PNG}c: $d_1=260$~nm, $d_2=140$~nm, $h=150$~nm.



\begin{figure}[h!]
\centering\includegraphics[width = .9\textwidth]{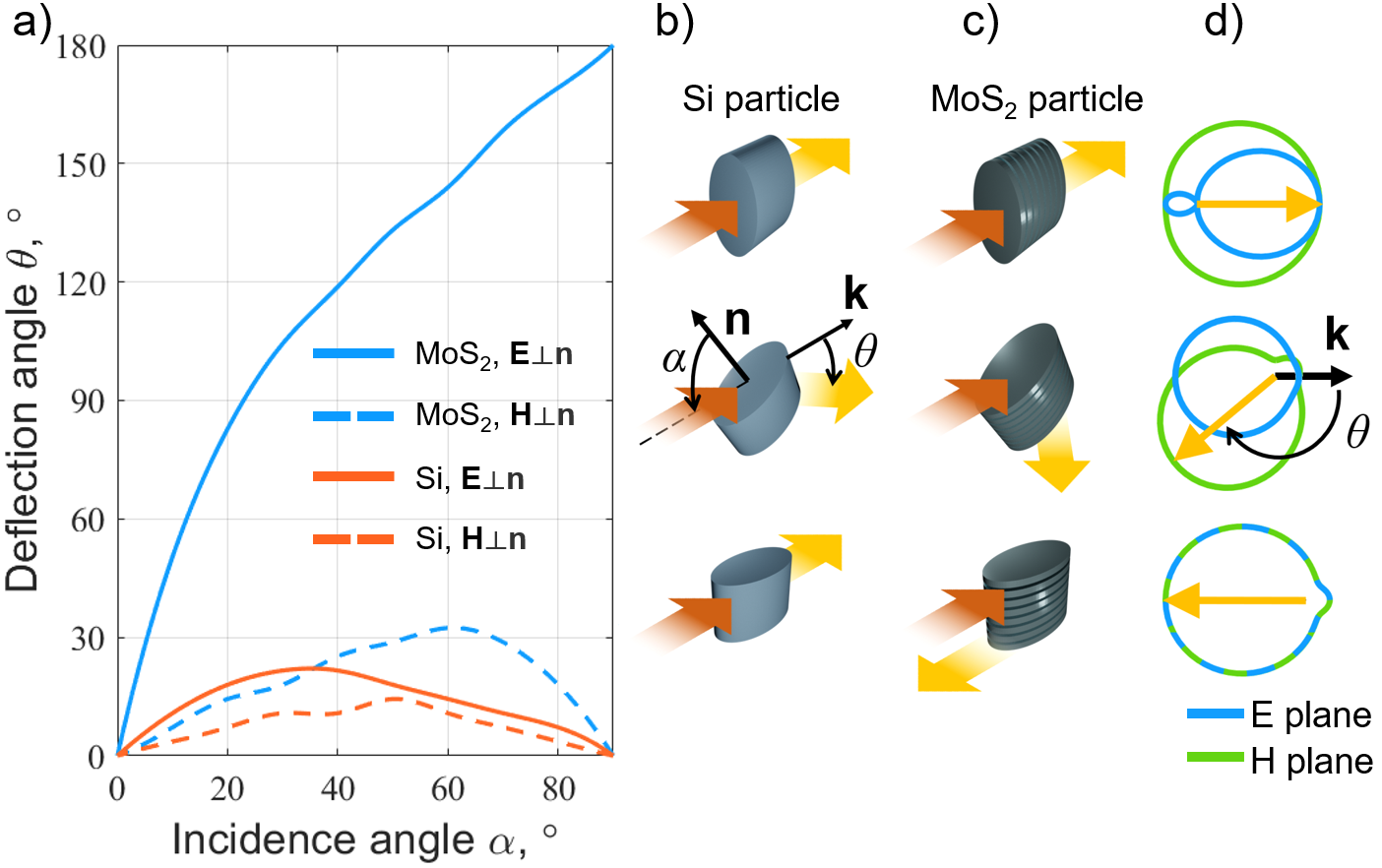}
\caption{Light deflection by MoS$_2$ and Si nanocylinders. The deflection is measured as an angle $\theta$ between the incident plane wave $\mathbf{k}$ and the direction of maximum scattered radiation in far field, see the definitions in b). (a) The deflection $\theta$ as a function of incidence angle $\alpha$ for MoS$_2$ and Si nanoparticles and two mutually orthogonal incident polarizations. The ``shape 2" particles with dimensions $d_1=260$~nm, $d_2=140$~nm, $h=150$~nm are considered (see Fig. \ref{MoS2andSilicon.PNG}a). b) and c) Schemes illustrating the differences in light deflection by Si and MoS$_2$ particles at different particles rotations. Vectors \textbf{n} and \textbf{k} in b) denote a normal vector to the particle plane and the incident wave vector, respectively; angles $\alpha$ and $\theta$ are defined on the same picture. The axis of particle rotation is normal to the plane with vectors \textbf{n}, \textbf{k} and the longest dimension $d_1$ of the particle (see also Fig. \ref{MoS2andSilicon.PNG}a); d) The cross-sections of the far field radiation pattern in E and H planes for the particles rotations shown in b) and c). The incident wavelength is 818~nm.}
\label{Deflection2.png}
\end{figure}

Figure \ref{Deflection2.png}a shows the efficiency of light deflection by Si and MoS$_2$ particles. We measure the efficiency of deflection as the angle $\theta$ between the incident plane wave $\mathbf{k}$ and the direction of maximum scattered radiation in far field, see Fig. \ref{Deflection2.png}b for definitions. 
Silicon particle deflects incident light by not more than $\sim$20$^\circ$, while the MoS$_2$ particle demonstrates the effect of ``super deflection" for TE polarization (with respect to normal vector \textbf{n}): 
the deflection angle varies continuously between $\theta=0^\circ$ (forward scattering) and $\theta=180^\circ$ (backward scattering) as the particle is being rotated between $\alpha=0^\circ$ and $\alpha=90^\circ$ incidence angles.
Although these results are shown only for a single wavelength of 818~nm, our simulations show that silicon ``shape 2" particles do not exhibit the super deflection regime at wavelengths 700~nm and longer, as well as MoS$_2$ ``shape 2" particles for TM polarization in the same wavelength region. 
The wavelength of 700~nm is the edge of the MoS$_2$ exciton band, see Fig. \ref{MoS2andSilicon.PNG}c. The MoS$_2$ particle, on the contrary, exhibits a ``super deflection" regime in TE polarization for wavelengths 750~nm $\lesssim\lambda\lesssim$ 850~nm.


\begin{figure*}[h!]
\includegraphics[width = \textwidth]{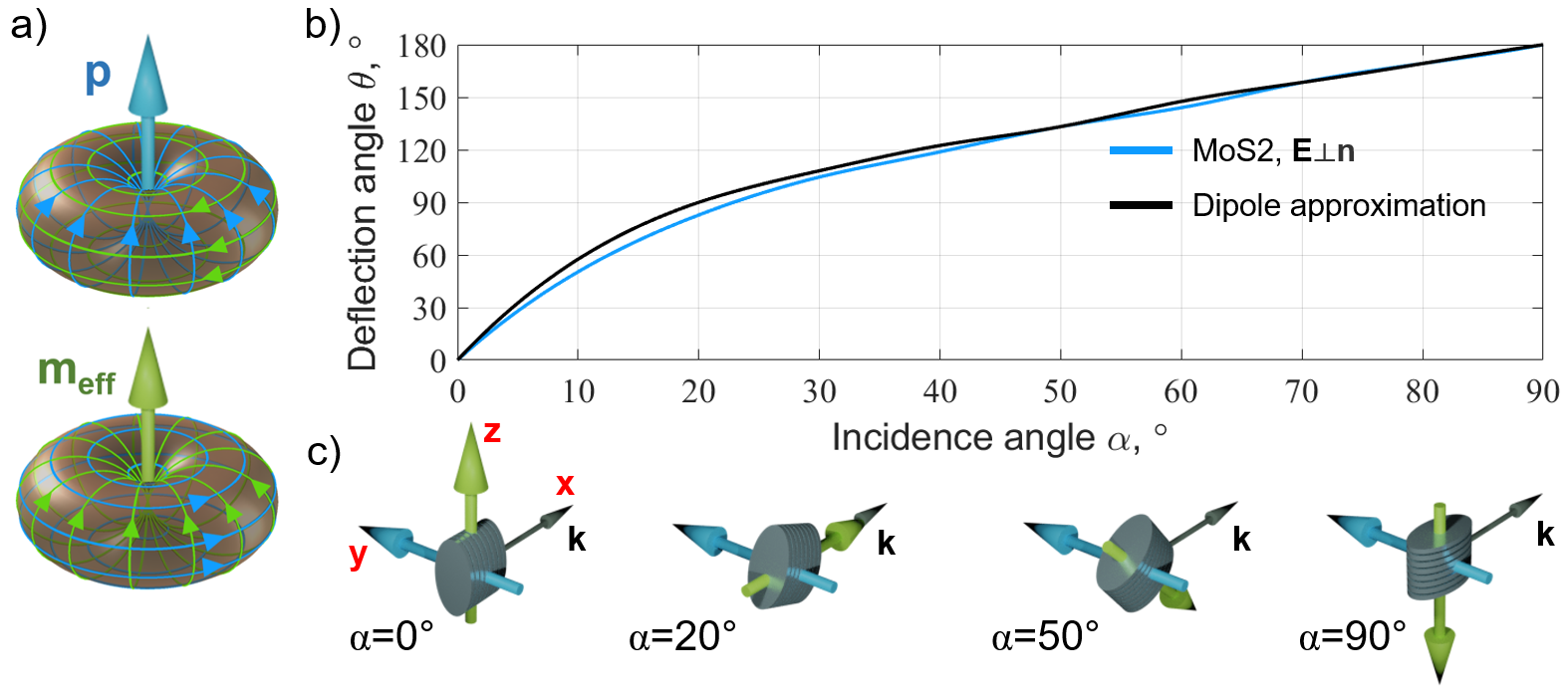}
\caption{Point dipole model of light deflection. (a) The electromagnetic polarization of electric (\textbf{p}) and magnetic (\textbf{m}) linear oscillating dipoles in the far field; b) blue curve: the rigorous calculations of the deflection angle $\theta$ as a function of incidence angle $\alpha$, and black curve is the same dependence obtained in the dipole approximation. c) Visualization of orientations of dipoles \textbf{p}, $\textbf{m}_{\text{eff}}$ and MoS$_2$ particle with respect to the incident plane wave with a wave vector \textbf{k}. Three axes $x,y,z$ in red denote a coordinate system, where the particle rotates.}
\label{DipoleApprox.PNG}
\end{figure*}

To elucidate the mechanism of this deflection, we describe light scattering by the MoS$_2$ nanoparticle using the dipole approximation with a number of simplifications. 
First, we only include electric and magnetic dipole moment contributions to the total scattered field.
Next, we simplify the induced electric and magnetic dipole moments.
Owing to the symmetry of the problem, the induced electric dipole moment \textbf{p} is always parallel to the $y$-axis, $\mathbf{p} \equiv p \hat{\mathbf{y}}$ (blue arrows in Fig. \ref{DipoleApprox.PNG}c).
The magnetic dipole moment \textbf{m}, in turn, always lies in the $xz$ plane and thus has two non-zero Cartesian components $m_z$ and $m_x$ (parallel to the wave vector \textbf{k}):
\begin{equation}
    \mathbf{m}\equiv\left( \begin{array}{c}
	m_x\\
	0\\
	m_z\\
\end{array} \right) =\left( \begin{array}{c}
	m_1\exp \left( i\varphi _1-i\omega t \right)\\
	0\\
	m_3\exp \left( i\varphi _3-i\omega t \right)\\
\end{array} \right),
\label{Eq4}
\end{equation}
These two Cartesian components have different amplitudes $m_1, m_3$ and phases $\varphi_1, \varphi_3$, so that the total magnetic dipole moment \textbf{m} is generally elliptically polarized in the $xz$ plane with longer and shorter ellipsoid axes
\begin{equation}
    \mathbf{m}_{\mathrm{long}}=\left( \begin{array}{c}
	m_1\cos \left( \varphi _1-\zeta /2 \right)\\
	0\\
	m_3\cos \left( \varphi _3-\zeta /2 \right)\\
\end{array} \right),\ \mathbf{m}_{\mathrm{short}}=\left( \begin{array}{c}
	m_1\sin \left( \zeta /2-\varphi _1 \right)\\
	0\\
	m_3\sin \left( \zeta /2-\varphi _3 \right)\\
\end{array} \right),
\label{Eq5}
\end{equation}
respectively. Here we utilized an auxiliary angle $\zeta$, defined via the following formulae:
\begin{equation}
    \begin{cases}
	\mathrm{A}=\frac{m_{1}^{2}}{2}\cos \left( 2\varphi _1 \right) +\frac{m_{3}^{2}}{2}\cos \left( 2\varphi _3 \right)\\
	\mathrm{B}=\frac{m_{1}^{2}}{2}\sin \left( 2\varphi _1 \right) +\frac{m_{3}^{2}}{2}\sin \left( 2\varphi _3 \right)\\
	\sin \left( \zeta \right) =\mathrm{B}/\sqrt{\mathrm{A}^2+\mathrm{B}^2}\\
	\cos \left( \zeta \right) =\mathrm{A}/\sqrt{\mathrm{A}^2+\mathrm{B}^2}\\
\end{cases}
\label{Eq6}
\end{equation}
We replace the elliptically oscillating magnetic dipole moment $\textbf{m}$ with a linearly polarized effective one $\textbf{m}_{\text{eff}}$, which is defined as a projection of the time-dependent vector $\textbf{m}(t)$ along the direction $\mathbf{m}_{\mathrm{long}}$. After some math we get a compact expression:
\begin{equation}
    \mathbf{m}_{\mathrm{eff}}=\mathbf{m}_{\mathrm{long}}\exp \left( i\zeta /2-i\omega t \right)
\label{Eq7}
\end{equation}
in Cartesian coordinates.

Figure \ref{DipoleApprox.PNG}b shows the results of applying the dipole approximation, which are in a good agreement with rigorous full-wave simulations involving higher-order multipoles. 
The performed two-step dipole approximation reveals the mechanism of the ``super deflector" effect: $\textbf{m}_{\text{eff}}$ rotates with a MoS$_2$ particle, whereas the direction of \textbf{p} stays unchanged (see Figs.\ref{DipoleApprox.PNG}c), and their interference changes the direction of scattering maximum. The presence of material anisotropy facilitates the rotation of $\textbf{m}_{\text{eff}}$: circular displacement currents ``feel" the layered nature of MoS$_2$, so the incident angle variations lead to noticeable redistribution of induced currents inside the particle. The ``super deflector" regime might, in principle, be found in Si particles too, by making its shape more asymmetric than in the considered case.

\section{Conclusion}

To conclude, we investigated both theoretically and numerically the scattering regimes of asymmetric particles made of silicon and MoS$_2$. We compared their optical response for the fixed geometries and found that the MoS$_2$ material anisotropy manifests itself through a number of features. Firstly, this anisotropy allows a fine tuning of anapole and magnetic dipole spectral positions and engineering the particles with structure-induced magnetic dipole scattering effect. Secondly, MoS$_2$ particle of the same geometry demonstrates prominent polarization-sensitive deflector properties in a certain frequency band. We have shown that the effect originates from the rotation of the magnetic dipole moment induced in the particle volume, which interferes with the electric dipole moment having a fixed direction during the particle rotations. Both effects, which are present simultaneously in a single particle, might be helpful for engineering dielectric beam-steering metasurfaces and stimulate experimental studies of anapole in non-conventional optical materials.


\begin{acknowledgments}

We gratefully acknowledge the financial support from the Ministry of Science and Higher Education of the Russian Federation (Agreement No. 075-15-2021-606). G.A.E. acknowledges the support from Russian Science Foundation (RSF) under grant number 22-29-01192. 
D.G.B. acknowledges support from Council on grants of the President of the Russian Federation (MK-1211.2021.1.2).
The multipole analysis  was  supported by  the Russian Science Foundation, Grant No. 20-12-00343. The authors acknowledge the support of the Azrieli Foundation's Postdoctoral Fellowship.
We gratefully acknowledge prof. Evlyukhin for fruitful discussions.

\end{acknowledgments}



\bibliography{TMDC}

\end{document}